\documentclass[pre,twocolumn,floatfix]{revtex4-1}

\newcommand{\expt}[1]{\left< #1\right>}
\newcommand{\Eq}[1]{Eq.~(\ref{eq:#1})}

\newcommand{\Fig}[1]{Fig.~\ref{fig:#1}}
\newcommand{\Figs}[1]{Figs.~\ref{fig:#1}}
\newcommand{\Ref}[1]{Ref.~\cite{#1}}
\newcommand{\Sec}[1]{Sec.~\ref{sec:#1}}
\newcommand{\Figure}[1]{Figure~\ref{fig:#1}}
\newcommand{\gdot}{\dot{\gamma}}

\newcommand{\gy}{g_\perp}

\newcommand{\Gy}{G_\perp}
\newcommand{\rr}{\mathbf{r}}
\newcommand{\rp}{\mathbf{r}'}
\newcommand{\skippa}[1]{}
\newcommand{\fast}{\mathrm{fast}}
\newcommand{\phiJ}{\phi_J}
\newcommand{\slow}{\mathrm{slow}}
\newcommand{\sym}{\mathrm{sym}}

\newcommand{\xhat}{\hat{x}}

\newcommand{\f}{\mathbf{f}}
\renewcommand{\v}{\mathbf{v}}
\newcommand{\Note}[1]{}

\usepackage{graphicx}

\begin{document}
\title{Asymmetric velocity correlations in shearing media}

\author{Peter Olsson}

\affiliation{Department of Physics, Ume\aa\ University, 
  901 87 Ume\aa, Sweden}

\date{\today}   

\begin{abstract}
  A model of soft frictionless disks in two dimensions at zero temperature is
  simulated with a shearing dynamics to study various kinds of asymmetries in
  sheared systems. We examine both single particle properties, the spatial
  velocity correlation function, and a correlation function designed to
  separate clockwise and counter-clockwise rotational fields from one
  another. Among the rich and interesting behaviors we find that the velocity
  correlation along the two different diagonals corresponding to compression and
  dilation, respectively, are almost identical and, furthermore, that a feature
  in one of the correlation functions is directly related to irreversible
  plastic events.
\end{abstract}


\maketitle

\section{Introduction}

In collections of particles with repulsive contact interaction there is a
transition from a liquid to an amorphous solid state as the volume fraction
increases --- the jamming transition. It has been suggested that this transition
is a critical phenomenon with universal critical exponents \cite{Liu_Nagel} and
the successful scaling of rheology data from simulations is strong evidence that
that actually is the case
\cite{Olsson_Teitel:jamming,Hatano:2008,Otsuki_Hayakawa:2009b}. The precise
values of the critical exponents, however, continue to be a matter of
discussion\cite{Tighe_WRvSvH}.

At the very heart of critical phenomena is the notion of a correlation length
that diverges as the critical point is approached. It is therefore important to
identify the proper correlation length. Several works have tried to look for a
growing order in the static quantities, but without much success. Another
possibility is to look for a growing length in the dynamics. Velocity
correlations in sheared systems were studied in \Ref{Ono_Tewari_Langer_Liu}
though not revealing any growing length. A large correlation length was however
found in \Ref{Lois_Lemaitre_Carlson} and they also argued for a pronounced
angular dependence of the velocity correlations \cite{Majmudar_Behringer:2005}.

In a previous work we reported the finding of a growing characteristic length
from the transverse component of the velocity correlation function
\cite{Olsson_Teitel:jamming}. The extraction of the correlation length exponent,
however, seems to be more complicated than presented there and we therefore set
out to do a more thorough analysis of the velocity correlations. As an important
step in that direction we here consider some symmetry properties of velocity
correlations in a sheared system and find a surprisingly rich and interesting
behavior

When shearing simulations are done slowly enough it becomes possible to separate
the time evolution into elastic parts where the energy slowly increases and
plastic contributions which are irreversible processes where the system rapidly
evolves and dissipates energy \cite{Maloney_Lemaitre:2004}. We will argue below
that the contribution from the plastic processes also may be seen in the
velocity correlation function.

The content of the present paper is the following: In \Sec{ModSim} we briefly
describe the model and the simulations.  \Sec{VelCorr} describes some rather
direct measures of velocity correlations and how they depend on the direction of
the separation between the particles whereas \Sec{Rot} deals with a more
involved correlation function designed to capture the difference between
clockwise and counter-clockwise rotations of the velocity field. A summary and
some concluding remarks are given in \Sec{Summ}.

\section{Model, simulations, and measured quantities}\label{sec:ModSim}

\subsection{Shearing dynamics}
Following O'Hern \emph{et al.}\ \cite{OHern_Silbert_Liu_Nagel:2003} we simulate
frictionless soft disks in two dimensions using a bi-dispersive mixture with
equal numbers of disks with two different radii of ratio 1.4. Length is measured
in units of the small particles ($d_s=1$). With $r_{ij}$ for the distance
between the centers of two particles and $d_{ij}$ the sum of their radii, the
interaction between overlapping particles is
\begin{displaymath}
  V(r_{ij}) = \left\{
    \begin{array}{ll}
      \frac{\epsilon}{2} (1 - r_{ij}/d_{ij})^2,\quad & r_{ij} < d_{ij},\\
      0, & r_{ij} \geq d_{ij}.
    \end{array} \right.
\end{displaymath}
We use Lees-Edwards boundary conditions \cite{Evans_Morriss} to introduce a
time-dependent shear strain $\gamma = t\gdot$. With periodic boundary conditions
on the coordinates $x_i$ and $y_i$ in an $L\times L$ system, the position of
particle $i$ in a box with strain $\gamma$ is defined as $\rr_i = (x_i+\gamma
y_i, y_i)$ which thus gives a shear flow in the $x$ direction. We simulate
overdamped dynamics at zero temperature with the equation of motion
\cite{Durian:1995},
\begin{displaymath}
  \frac{d\rr_i}{dt} = -{C}\sum_j\frac{dV(\rr_{ij})}{d\rr_i} + y_i \gdot\; \xhat,
\end{displaymath}
which is integrated with the second order Heuns' method.

This above expression is for the total velocity, including the shearing part. In
the analyzes of the velocity correlations below we will use the non-affine part
of the velocity excluding the trivial shearing part $y_i\gdot\xhat$. This
non-affine part of the velocity will be denoted by $\v$.

Our simulations are performed with $N=65536$ with shear rates down to
$\gdot=10^{-8}$. The averages are typically from simulations during 1--3 days
and nights with 128 cores on a massively parallel computer.

\section{Velocity correlations}\label{sec:VelCorr}

\subsection{Symmetry of a shearing system}

\begin{figure}
  \includegraphics[width=6cm]{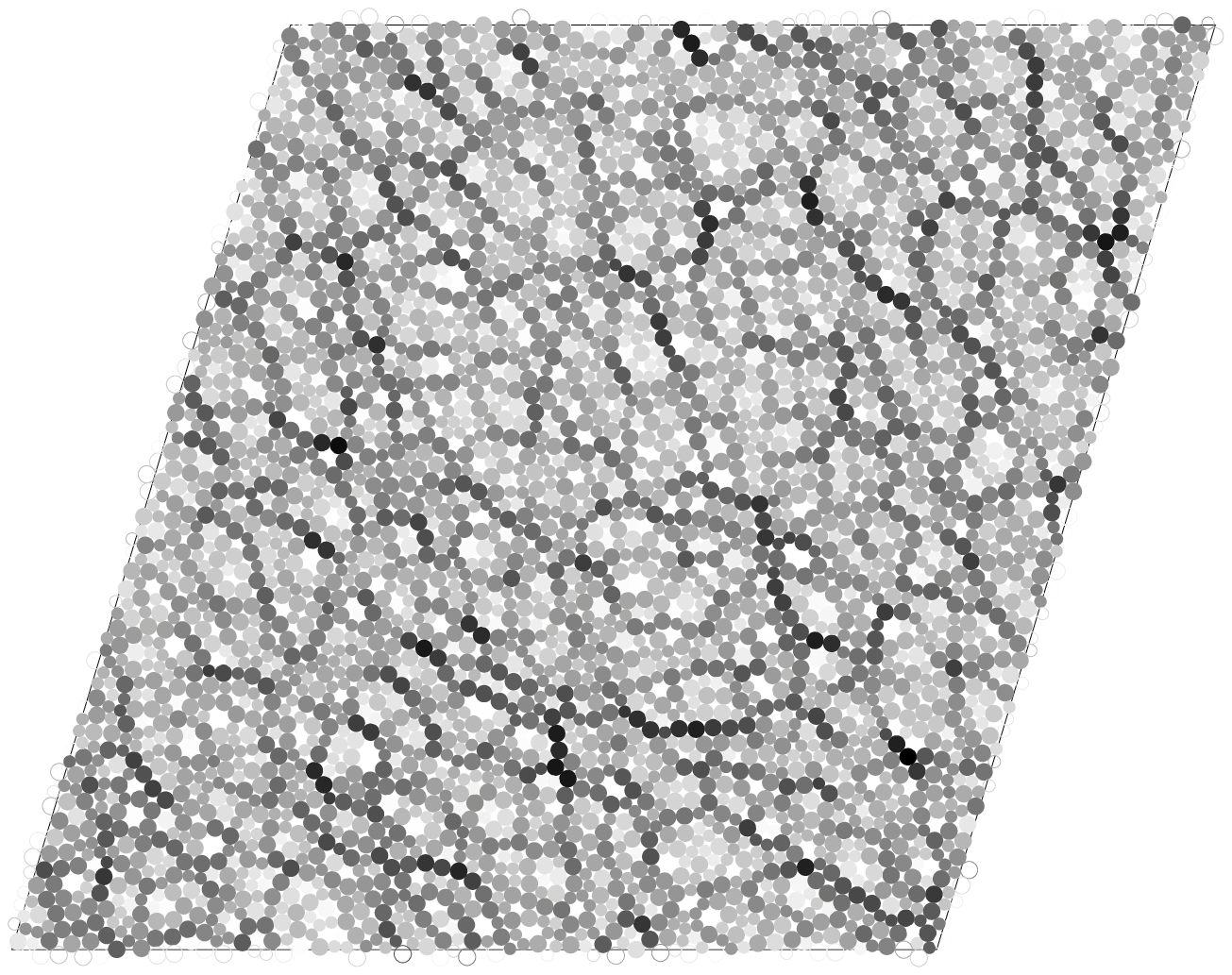}
  \caption{Configuration with 4096 particles color coded according to the
    elastic energy of each particle. The dark particles (high elastic energy)
    make up force chains that preferably extend along the $\hat x - \hat y$
    direction.}
  \label{fig:vlat}
\end{figure}

\Figure{vlat} shows the presence of force chains in our system. The figure is a
configuration with 4096 particles, color coded according to the elastic energy
of each particle. Note that the force chains\cite{Majmudar_Behringer:2005} tend
to be along the $\hat x - \hat y$ direction, which is the direction of
compression.  This means that the force chains break the reflection symmetry
along $x$ and that the system is only symmetric under the \emph{combined}
transformation $x\rightarrow -x$ and $y\rightarrow -y$. The same conclusion is
readily drawn from the expression for the shear stress,
\begin{displaymath}
  \sigma = \frac{1}{L^2} \sum_{ij} f^x_{ij} y_{ij},
\end{displaymath}
where $L$ is the linear system size, the sum is over pairs of particles,
$f^x_{ij}$ is the $x$-component of the force between particles $i$ and $j$, and
$y_{ij}$ is the $y$-component of their separation. Note that the shear stress
changes sign under the transformation $x\rightarrow -x$ but remains unchanged
under the combined transformation $x\rightarrow -x$ and $y\rightarrow -y$.

\subsection{Single particle properties}\label{sec:Basic}

As we will see below several symmetries that hold in systems at equilibrium are
broken in a shearing system. The simplest symmetry is however respected; there
is no net velocity in the system, $\expt{\v} = 0$. As remarked above, $\v$ is
the non-affine part of the velocity. Here and in the following $\expt\ldots$
represents the average over all particles and a large number of configurations
generated with our shearing dynamics. For the average velocity things are
unusually simple since the same result holds for each individual configuration
as a consequence of the overdamped dynamics and total force balance, $\v =
\sum_i \v_i = C\sum_{ij} \f_{ij} = 0$ as $\f_{ij} = -\f_{ji}$.

The conclusion of a vanishing average velocity may also be reached from the
symmetry considerations.  Since the combined transformation also implies the
change of sign of both velocity components, $v_x \rightarrow -v_x$ and $v_y
\rightarrow -v_y$, it follows that $\expt{v_\mu} = \expt{-v_\mu}$, for
$\mu=x,y$, which gives $\expt{v_\mu} = 0$.

\begin{figure}
  \includegraphics[width=8cm]{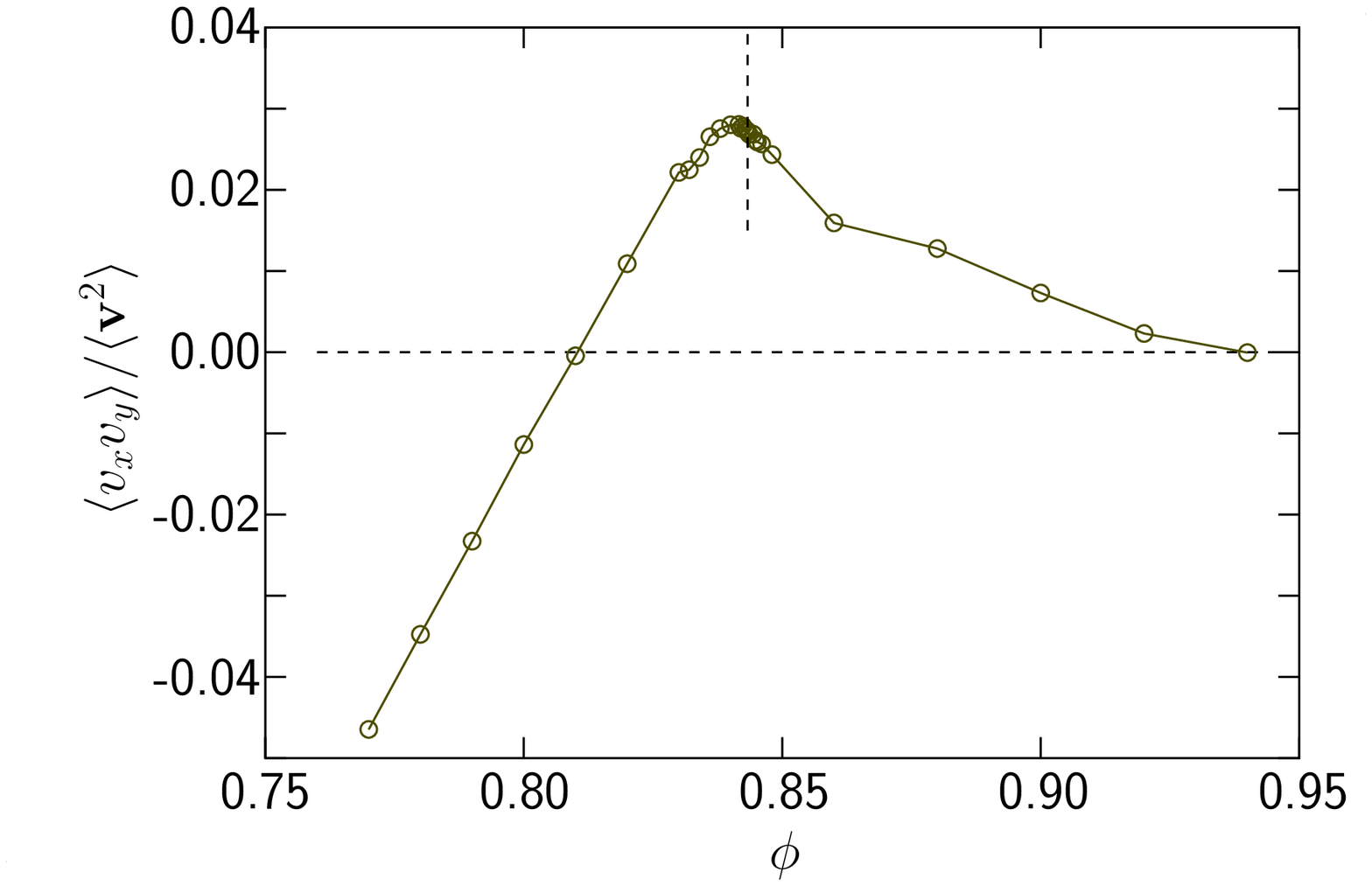}
  \includegraphics[width=8cm]{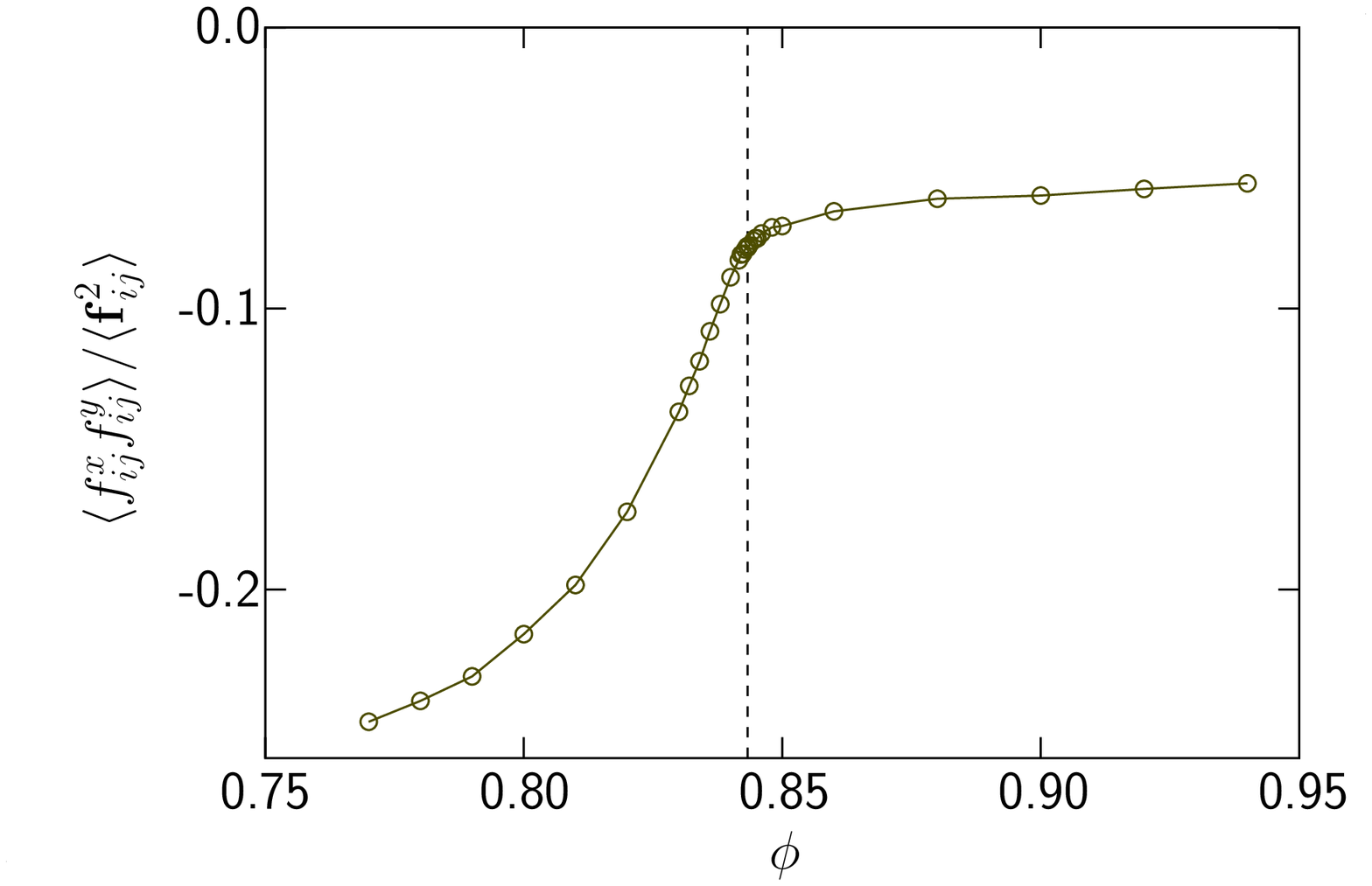}
  \caption{Measure of the anisotropy in both particle velocities and contact
    forces from simulations with $\gdot=10^{-7}$.  Panel (a) shows the
    normalized $\expt{v_x v_y}$ for individual particles versus density. The
    correlation changes from negative at low densities to positive around
    $\phi\approx \phiJ$ (dashed vertical line), signifying a change from a
    predominance of motion along the force chains to a slight overweight for
    motion perpendicular to the force chains. Panel (b) shows that the
    corresponding contact force correlation is always negative, which is
    consistence with the existence of force chains at all densities. This
    contact force correlation also has a marked feature at $\phi\approx
    \phiJ$. }
  \label{fig:vxvy}
\end{figure}

In contrast to equilibrium results from symmetry that $\expt{v_x v_y}=0$, one
finds that this quantity does not vanish in the sheared system. \Fig{vxvy} shows
$\expt{v_x v_y}/\expt{\v^2}$ against $\phi$.  At low densities the correlation
is negative which means that the particles tend to move more along than
perpendicular to the force chains. The correlation changes sign at
$\phi\approx0.81$, reaches a peak at $\phi\approx 0.84\approx\phiJ$ and then
decrease towards zero. This means that there is a region around $\phi_J$ where
the particles move slightly more in the direction \emph{perpendicular to} the
force chains.

A related quantity is $\expt{f_{ij}^x f_{ij}^y}$ which is the correlation
between the different components of the contact force $\f_{ij}$. Note that there
is a direct relation between the velocities and the contact forces: $\v_i =
C\sum_j \f_{ij}$, where the sum extends over all particles $j$ in contact with
$i$.  Nevertheless, $\expt{f_{ij}^x f_{ij}^y}$ behaves rather differently from
the velocity correlations.  Whereas $\expt{v_x v_y}/\expt{\v^2}$ has a peak at
$\phi\approx\phiJ$, \Fig{vxvy}(b) shows that the normalized $\expt{f_{ij}^x
  f_{ij}^y}$ changes at the same density from a rapid increase (which is a
\emph{decrease} in magnitude) to being almost constant.

\begin{figure}
  \includegraphics[width=8cm]{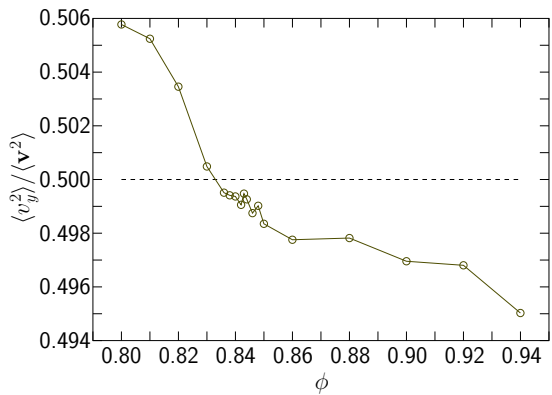}
  \caption{The anisotropy measured through the $\expt{v_y^2}$ relative to
    $\expt{\v^2} = \expt{v_x^2} +\expt{v_y^2}$. This fraction is close to 50\%
    in the regions around jamming but decreases slightly with increasing
    density.}
  \label{fig:vy2-phi}
\end{figure}

Another measure of the asymmetry is the relative magnitude of the two velocity
components. From energy balance---that the dissipated power has to be equal to
the supplied power---follows the result for the velocity squared: $C\expt{\v^2}
= (L^2/N) \sigma\gdot$ \cite{Ono_Tewari_Langer_Liu}. This dissipated power needs
however not be equal in the $x$ and $y$ directions. \Fig{vy2-phi} shows that the
fraction of the power dissipated by velocities along the $y$ direction is close
to 50\%, but also that there is a clear dependence on density:
$\expt{v_y^2}/\expt{\v^2}$ decreases from 0.506 to 0.495 when the density
increases from $\phi=0.80$ to 0.94.

\subsection{Spatial dependence}

We now turn to the spatial velocity correlations, i.e.\ the correlations between
pairs of particles with separation $\rr$. We first focus on the total
correlation function,
\begin{equation}
  \label{eq:gcorr}
  g(\rr) = \expt{\v(\rp) \cdot \v(\rp+\rr)},
\end{equation}
and examine how it depends on both magnitude $r=|\rr|$ and direction of $\rr$.
We will study this correlation function in four different directions: along the
two main directions, $\hat x$, $\hat y$, and along the diagonals
\begin{eqnarray*}
  \hat s & = & (\hat x + \hat y) / \sqrt 2, \\
  \hat t & = & (\hat y - \hat x) / \sqrt 2.
\end{eqnarray*}
In our shearing geometry $\hat t$ is the direction of uniaxial compression and
$\hat s$ the direction of uniaxial dilation.
\begin{figure}
  \includegraphics[width=8cm]{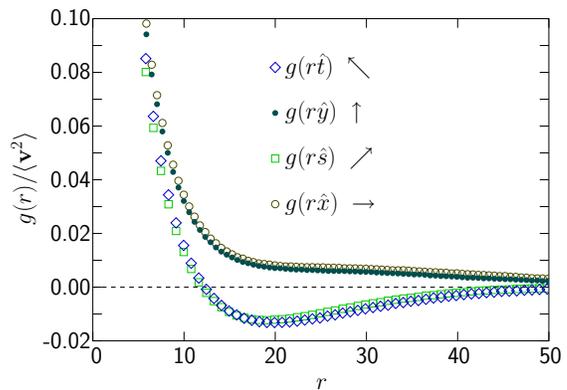}
  \caption{Total velocity correlation along the four different main
    directions. The decay of the correlations is monotonic along the main
    directions, $\hat x$ and $\hat y$ and non-monotonic along the diagonals.}
  \label{fig:gcorr}
\end{figure}
\Figure{gcorr} shows the correlation functions along these four different
directions. The curves are pairwise equal with, on the one hand, the
correlations along the main directions, $\hat x$ and $\hat y$, and on the other
hand the correlations along the diagonals. These results should be essentially
without finite size effects since the system size is $L\approx300$ whereas $r$
only extends up to $50$.  In view of the density dependence in \Figs{vxvy} and
\ref{fig:vy2-phi} we have confirmed that the general behavior remains the same
for a wide range of densities around $\phiJ$.

The result that $g(\rr)$ behaves the same along both diagonals is different from
the earlier finding of an angular dependence of the correlation length in
shearing systems \cite{Lois_Lemaitre_Carlson} with minimum and maximum along the
different diagonals. The reason for this difference is not clear, but we
speculate that it is related to the very different dynamics in their system,
which is also reflected in the oscillatory behavior of their velocity
correlation function.

\begin{figure}
  \includegraphics[width=8cm]{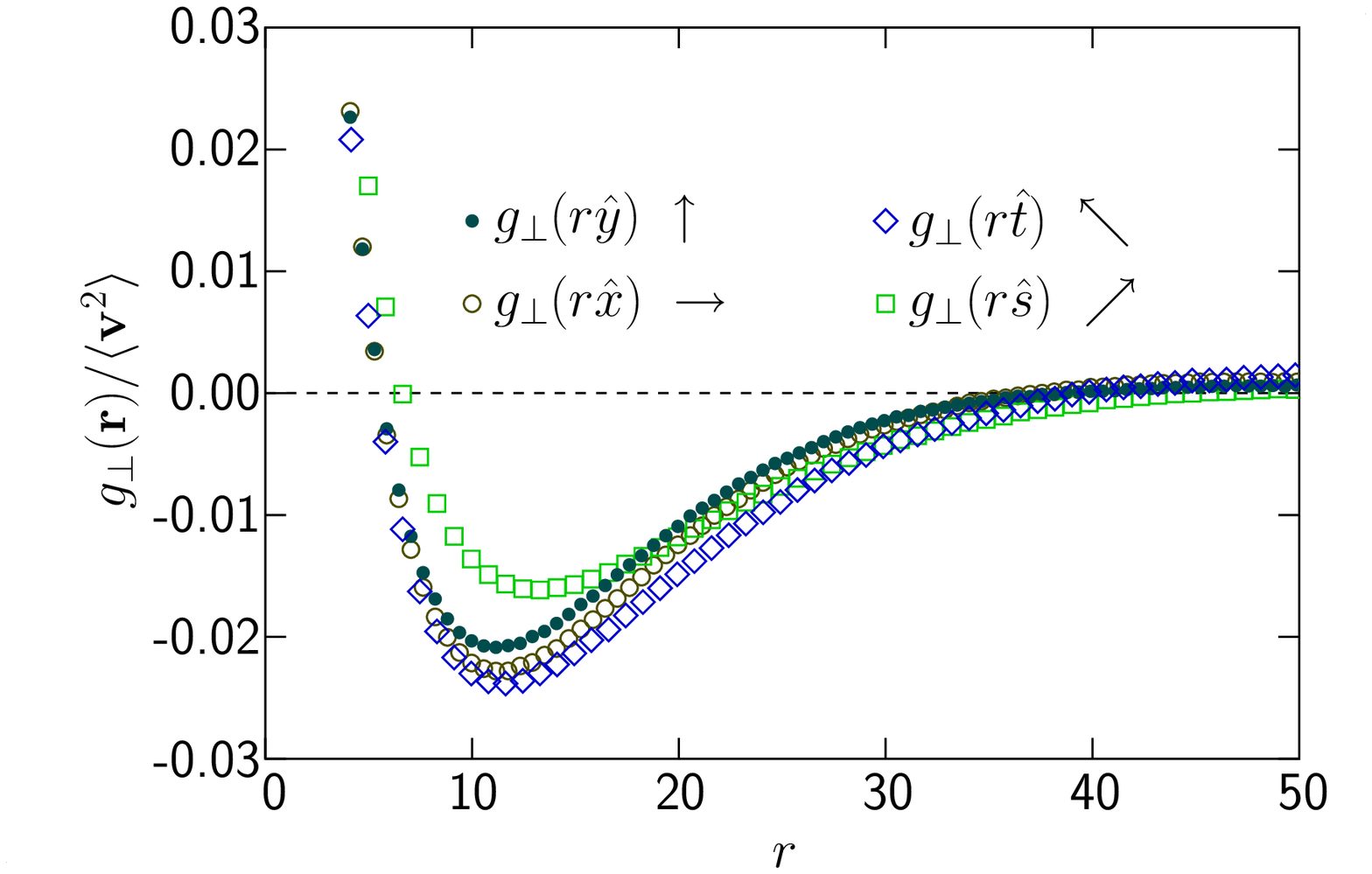}
  \includegraphics[width=8cm]{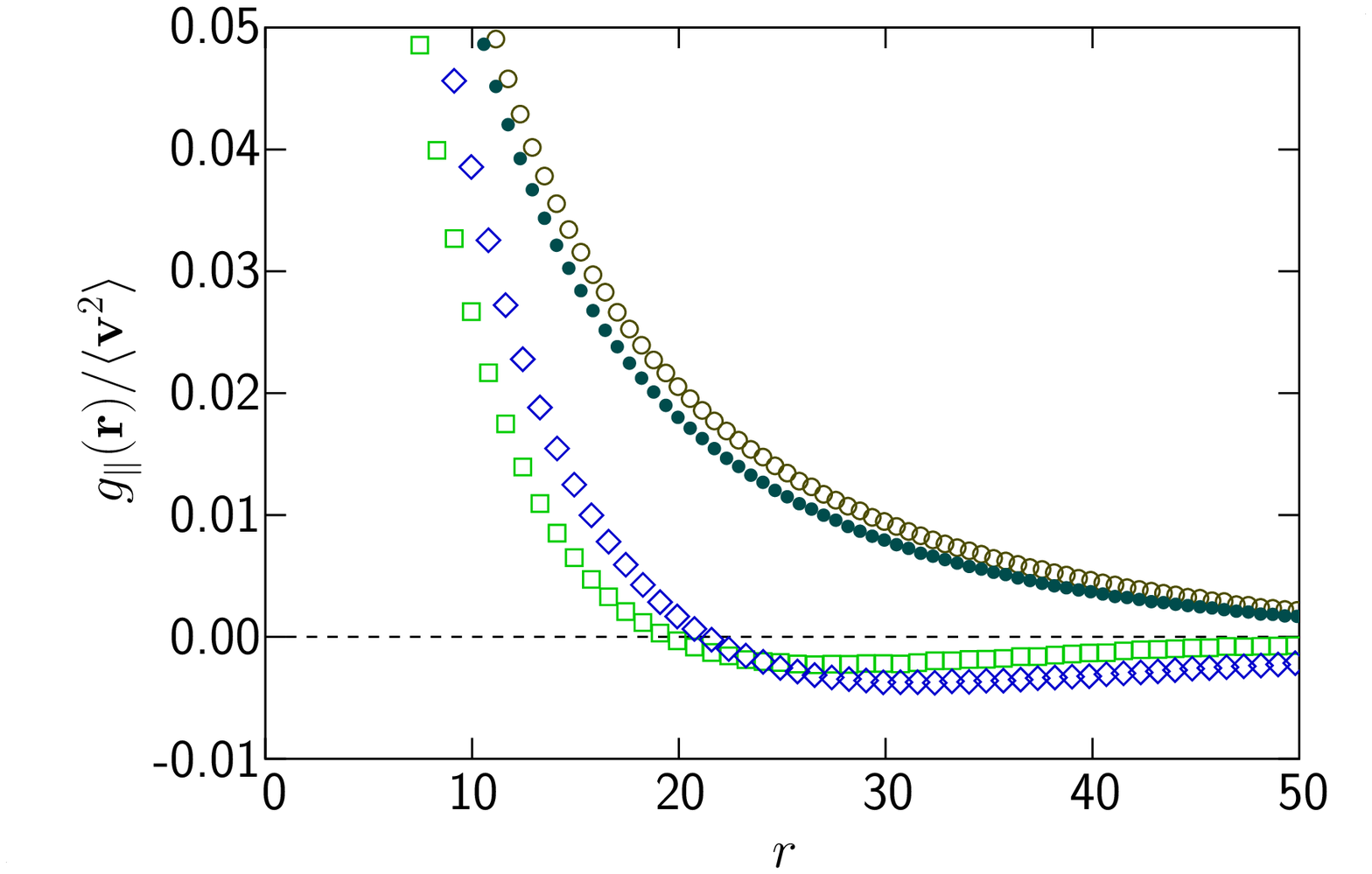}
  \caption{Transverse and longitudinal velocity correlations, respectively,
    along the four different directions. The transverse correlation in panel (a)
    is very similar for all four directions whereas the longitudinal correlation
    function shown in panel (b) is different for the diagonal directions,
    compared to the two main directions.  The differences in the total velocity
    correlations in \Fig{gcorr} are thus related to differences in the
    longitudinal correlations.}
  \label{fig:separate}
\end{figure}

To investigate the reason for the dependence of $g(\rr)$ on the direction of
$\rr$, we separate the correlations into longitudinal and transverse components,
parallel and perpendicular to the separation, respectively. In the $x$ direction
these components are
\begin{eqnarray*}
  g_\parallel(r\hat x) & = & \expt{v_x(0) v_x(r\hat x)}, \\
  g_\perp(r\hat x) & = & \expt{v_y(0) v_y(r\hat x)},
\end{eqnarray*}
with obvious generalizations to the other directions.  After this separation we
find that the transverse component along these four different directions behave
about the same, see \Fig{separate}(a). (The rather small differences in the
transverse component for the two diagonal directions will be discussed further
below.) The difference is largely due to the decay of the longitudinal
component, see \Fig{separate}(b) which is monotonic along the main ($\hat x$ and
$\hat y$) directions but non-monotonic along the diagonals.

The result above is a different behavior along the diagonals compared to the
main directions. We now instead focus on the difference between the two diagonal
directions, $\hat s$ and $\hat t$. From \Fig{gcorr} we found that the total
velocity correlations along these two directions are very similar. Nevertheless.
as shown in \Fig{separate} both $g_\parallel$ and $g_\perp$ are clearly
different. This suggests that this difference originates from the mixed
correlations $g_{xy}(\rr) = \expt{v_x(0) v_y(\rr)}$. To see this we use the
definitions $g_\parallel(r\hat s) = \expt{v_s(0) v_s(r\hat s)}$, and
$g_\perp(r\hat s) = \expt{v_t(0) v_t(r\hat s)}$ together with $v_s = (v_x +
v_y)/\sqrt 2$ and $v_t = (v_y - v_x)/\sqrt 2$ for the velocities along the
diagonals. This gives
\begin{eqnarray*}
  \label{eq:g_s}
  g_\parallel(r\hat s) & = & g(r\hat s) + g_{yx}(r\hat s), \\
  g_\perp(r\hat s) & = & g(r\hat s) - g_{yx}(r\hat s),
\end{eqnarray*}
where we have also made use of the symmetry $g_{xy} = g_{yx}$ which follows from
considering the transformation $\rr\rightarrow -\rr$, followed by a translation:
$v_x(0) v_y(\rr) \rightarrow (-v_x(0)) (-v_y(-\rr)) = v_x(\rr) v_y(0) = v_y(0)
v_x(\rr)$.

\begin{figure}
  \includegraphics[width=8cm]{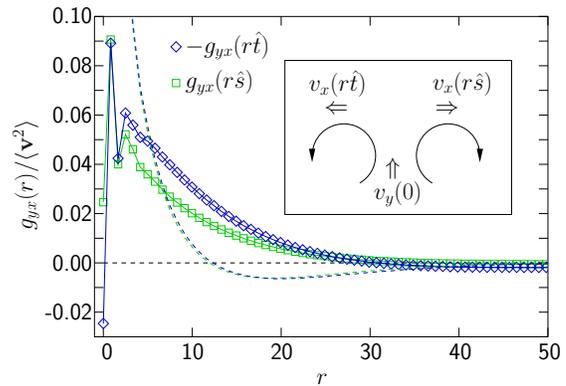}
  \caption{The mixed correlation functions $g_{yx}(r\hat s)$ and $-g_{yx}(r\hat
    t)$. The difference between the mixed correlation functions (apart from the
    trivially different sign) is yet an example of the asymmetry in the sheared
    system.  The arrows in the inset illustrate the rotational velocity fields
    which have the effect that a velocity $v_y(0)>0$ at the origin on the
    average gives $v_x(r\hat s)>0$ and $v_x(r\hat t)<0$.}
  \label{fig:Gyx-r}
\end{figure}
\Figure{Gyx-r} shows the mixed correlations along the two diagonal directions.
The thick arrows in the inset illustrate the velocity fields one might expect as
an effect of a particle with $v_y>0$ which is $v_x>0$ for separations in the
$\hat s$ direction and $v_x<0$ in the $\hat t$ direction. Since we expect
$v_x<0$ along $\hat t$, the correlations in that direction are shown with the
opposite sign. Note that both functions start out at $g_{yx}(0) = \expt{v_x
  v_y}$. The correlations then grow above this $r=0$ value and actually become
stronger in the $\hat t$ direction (though with opposite sign) than along $\hat
s$. For a comparison, the dashed lines are the correlations of $v_y$ along the
same directions.  This difference between $g_{yx}(r\hat s)$ and $-g_{yx}(r\hat
t)$ is yet another example of a broken symmetry in the shearing system.

\section{Rotational asymmetry}\label{sec:Rot}

The previous section gave evidence for the importance of whirls in the velocity
field. The inset of \Fig{Gyx-r} shows two such whirls that are a consequence of
$v_y(0)>0$, and there will also be whirls with the opposite orientation that
will contribute to the correlation functions in much the same way.  The question
we now like to address is whether they contribute equally much or not. Is the
system symmetric when considering whirls with clockwise and counter-clockwise
rotations, respectively?

Since the shearing by itself introduces a rotational field, it could at first
seem obvious that this symmetry is broken, but that is not correct. Our velocity
correlations are calculated from the \emph{non-affine} velocities and the net
rotation is zero in the non-affine velocity field. The system could therefore in
principle well be symmetric with respect to these different directions of
rotation.

\subsection{Asymmetric correlations}

There are at least two different ways to motivate the new correlation function
that we are about to introduce. The first is to note that the direction of the
rotational fields in the inset of \Fig{Gyx-r} depend on the sign of $v_y(0)$.
With the opposite sign of $v_y(0)$, $v_x(r\hat t)$ and $v_x(r\hat s)$ would also
(typically) change sign and the rotations would be in the opposite directions.

A second point of departure is to consider the fact that the correlation
function defined in \Eq{gcorr} is symmetric under the transformation
$x\rightarrow-x$ whereas the system itself is only symmetric under the
interchange of both $x$ and $y$. This suggests that some information is lost
when calculating the standard correlation function and, furthermore, that a
guiding principle in the definition of an ideal correlation function is that it
should have the same symmetry as the system.

It is then possible to combine both these lines of thought and construct a
function by restricting the average in \Eq{gcorr} to only include terms with
$v_y(\rp)>0$. For the transverse correlation function with $\rr$ along the $\hat
x$ direction, this becomes
\begin{equation}
  \label{eq:gplus}
  g_\perp^+(x) = \expt{v_y(\rp) v_y(\rp+x\hat x)}_{v_y(\rp)>0}.
\end{equation}
This expression may further be generalized to including particles with both
signs of $v_y(\rr)$ by letting the direction of the separation ($+x\hat{x}$ or
$-x\hat{x}$) depend on the sign of the velocity,
\begin{equation}
  \label{eq:gy-x}
  \gy^+(x) = \frac{1}{N_\mathrm{term}} \sum_{\rp}\left\{
    \begin{array}{ll}
      v_y(\rp) v_y(\rp+x\xhat), & v_y(\rp) > 0,\\
      v_y(\rp) v_y(\rp-x\xhat), & v_y(\rp) < 0.
    \end{array}
\right.
\end{equation}
The normalization $N_\mathrm{term}$ is the number of terms in the sum. Written
this way it becomes clear that $\gy^+(x)$ indeed has the desired symmetry
properties.  The rational for this new asymmetric function is also discussed in
conjunction with \Fig{Gy-fast} below. Note also that the \emph{symmetric}
function is related to $\gy(x)$ through
\begin{displaymath}
  \gy(x) = \frac{1}{2}[\gy^+(x) + \gy^+(-x)].
\end{displaymath}
The quantities shown in the figures below (both symmetric and asymmetric) are,
$G(x) = g(x)/g(0)$, normalized such that $G(0)=1$.
\begin{figure}
  \includegraphics[width=8cm]{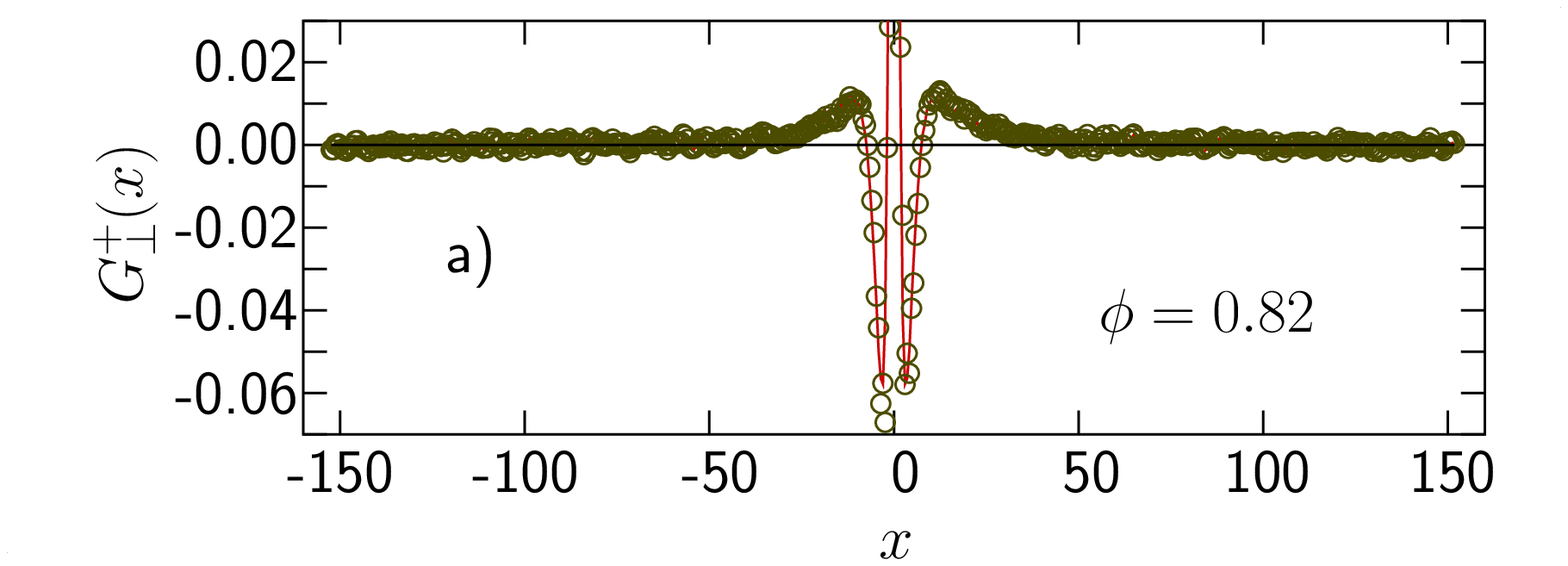}
  \includegraphics[width=8cm]{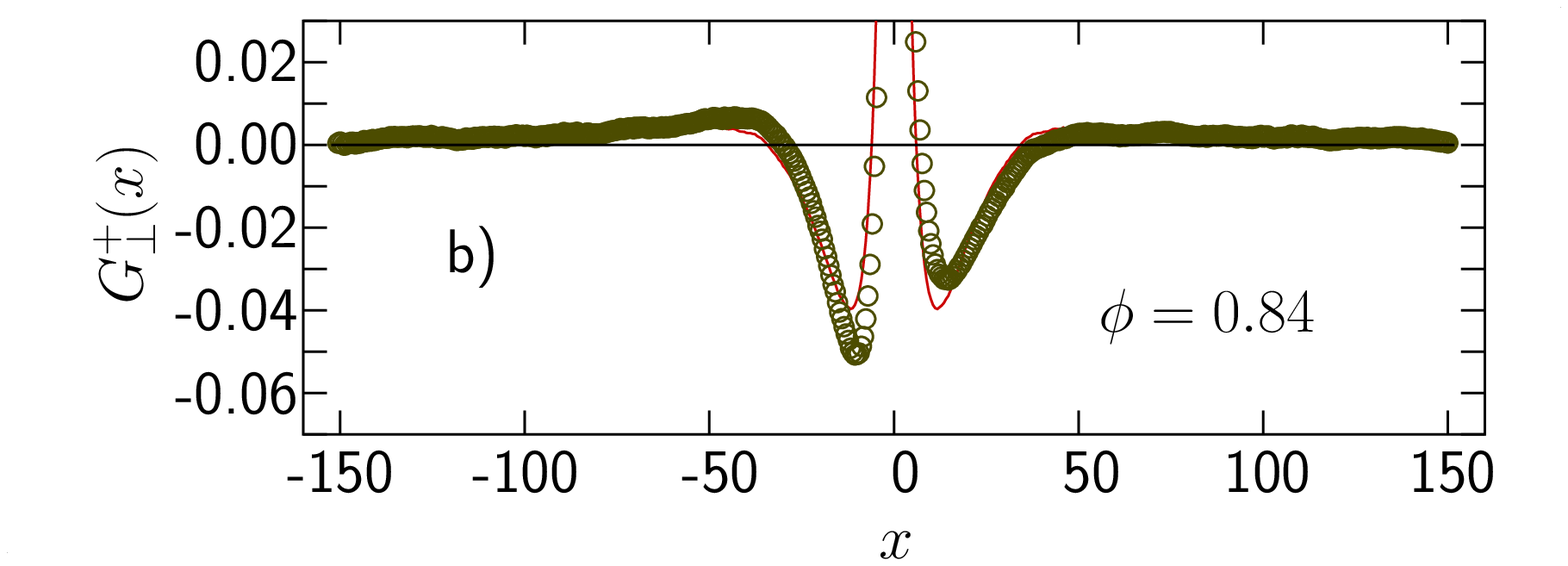}
  \includegraphics[width=8cm]{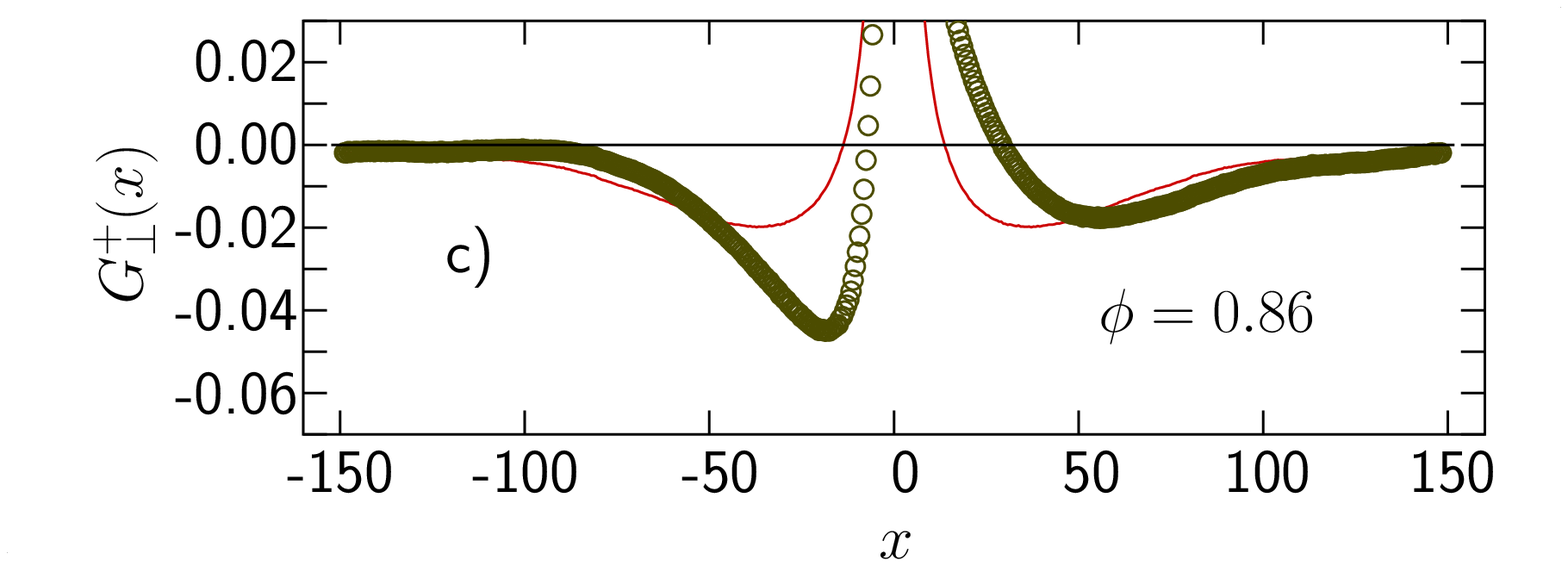}
  \includegraphics[width=8cm]{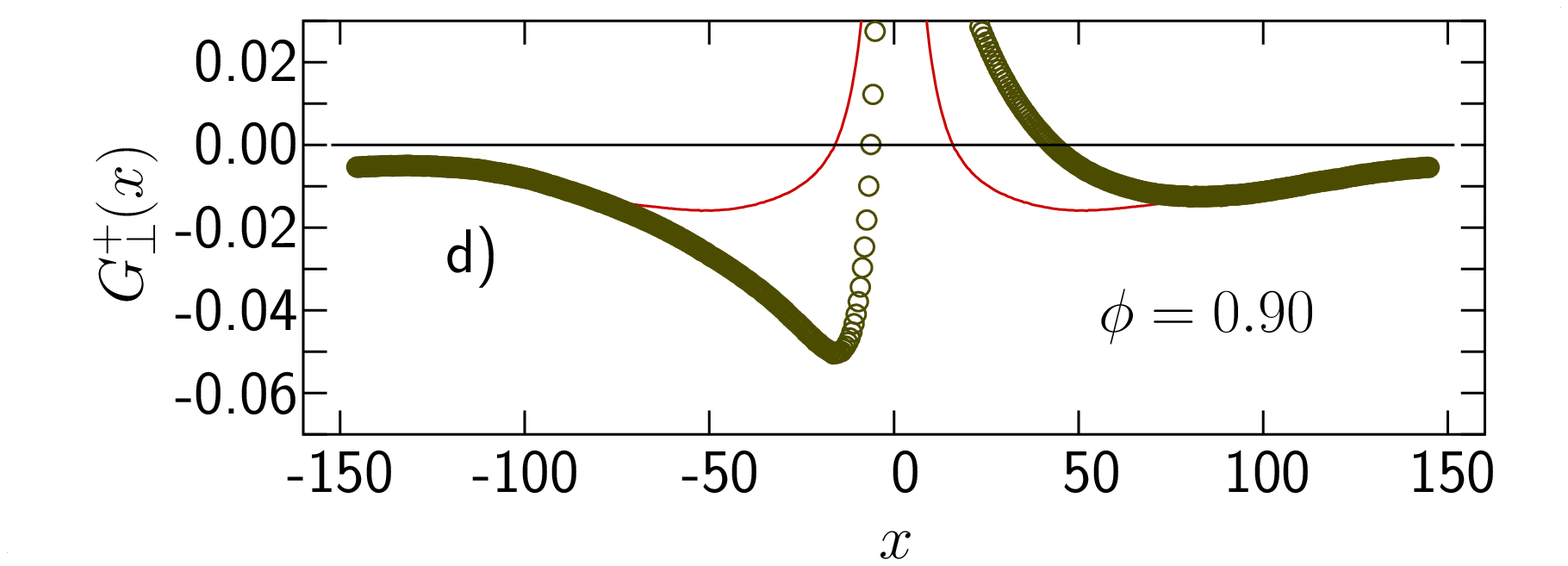}
  \caption{Development of the asymmetry in $\Gy^+(x)$. The function is nearly
    symmetric at $\phi=0.82$ but develops a very pronounced asymmetry as the
    density is increased above $\phiJ$. The solid lines are the symmetric
    $\Gy$.  The shear rate is $\gdot=10^{-7}$.}
  \label{fig:Gy-vi0001}
\end{figure}

\Figure{Gy-vi0001} shows $\Gy^+(x)$ at four different densities both above and
below $\phiJ$, again from simulations with $\gdot=10^{-7}$. For comparison, the
symmetrized function $\Gy(x)$ is given by the solid line. At $\phi=0.82$, panel
(a), $\Gy^+(x)$ is almost symmetric. The other panels show that the asymmetry
grows with increasing $\phi$, and at $\phi=0.90$ well above $\phiJ$, panel (d),
the asymmetry is very pronounced. The $x<0$ part shows a sharp dip at
$x\approx-18$ whereas the $x>0$ part has a rather shallow minimum at
$x\approx79$.

\begin{figure}
  \includegraphics[width=8cm]{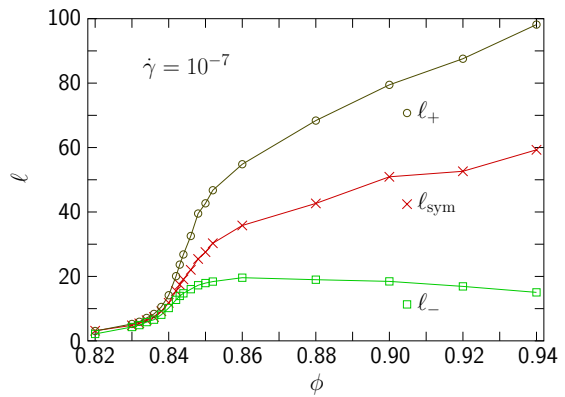}
  \caption{Location of the respective minima of the correlation
    functions. $\ell_+$ and $\ell_-$ are the positions of the minima of $\Gy^+$
    at $x>0$ and $x<0$, respectively, whereas $\ell_\sym$ is from the minimum of
    $\Gy$. The asymmetry of $\Gy^+$ which is reflected in the difference
    $\ell_+-\ell_-$ develops around $\phiJ$.}
  \label{fig:ell}
\end{figure}

Another way to illustrate the growth of the asymmetry is through the position of
the minima. \Fig{ell} shows $\ell_+$ and $\ell_-$, the absolute value of the
position of the minima for $x>0$ and $x<0$, respectively, together with
$\ell_\sym$ from the minimum of the symmetrized function. The asymmetry grows
rapidly above $\phi=0.84$ and the behavior of $\ell_+$ and $\ell_-$ turn out to
be very different at higher densities; $\ell_+$ continues to increase whereas
$\ell_-$ reaches a maximum at $\phi\approx 0.86$ and then decreases slowly.

\begin{figure}
  \includegraphics[width=8cm]{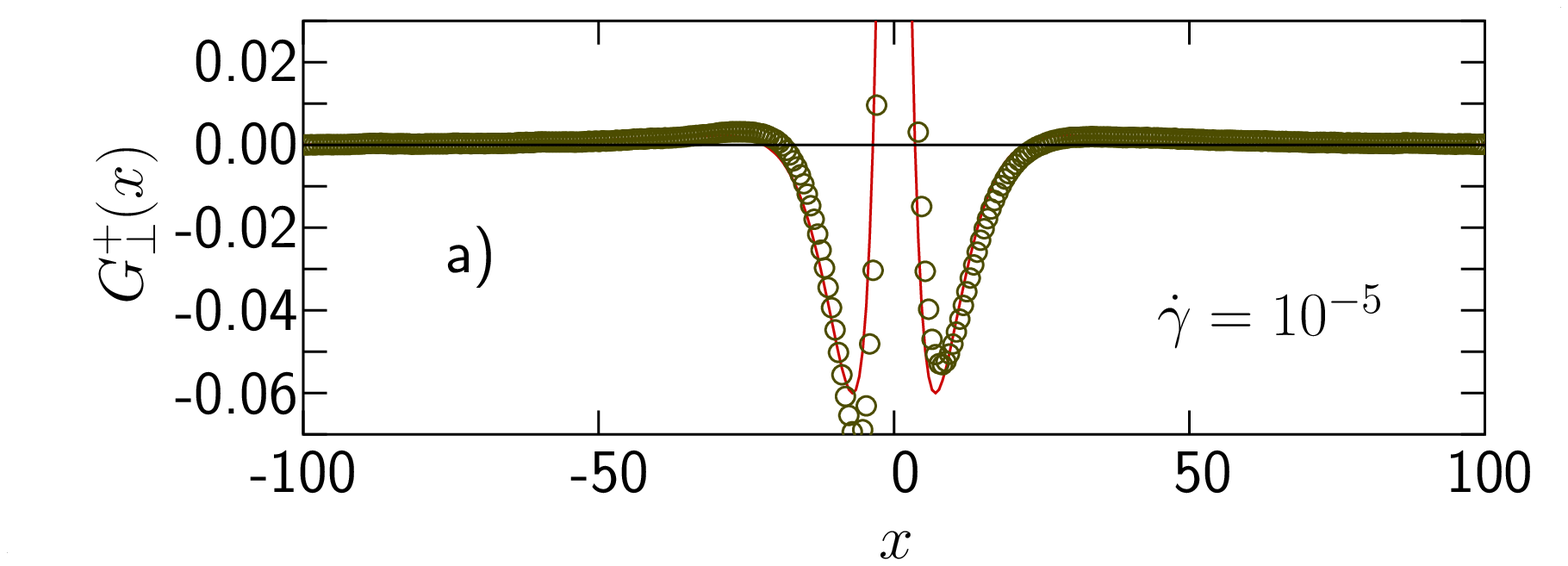}
  \includegraphics[width=8cm]{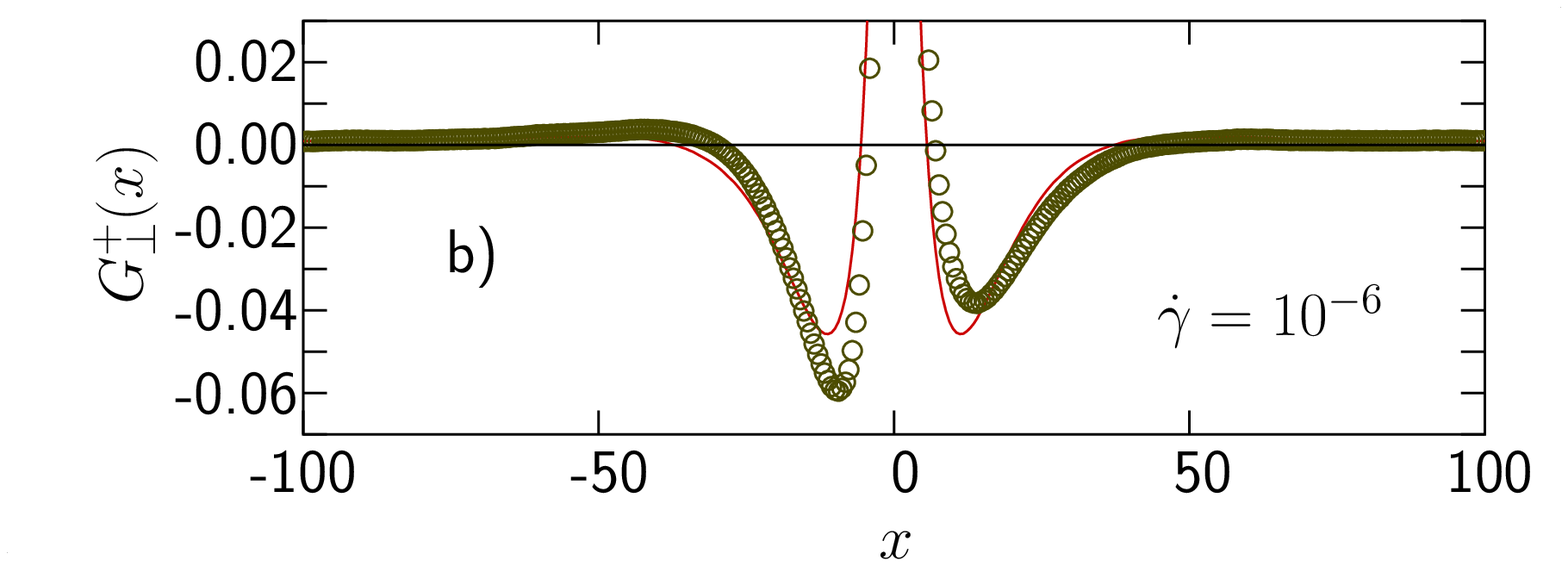}
  \includegraphics[width=8cm]{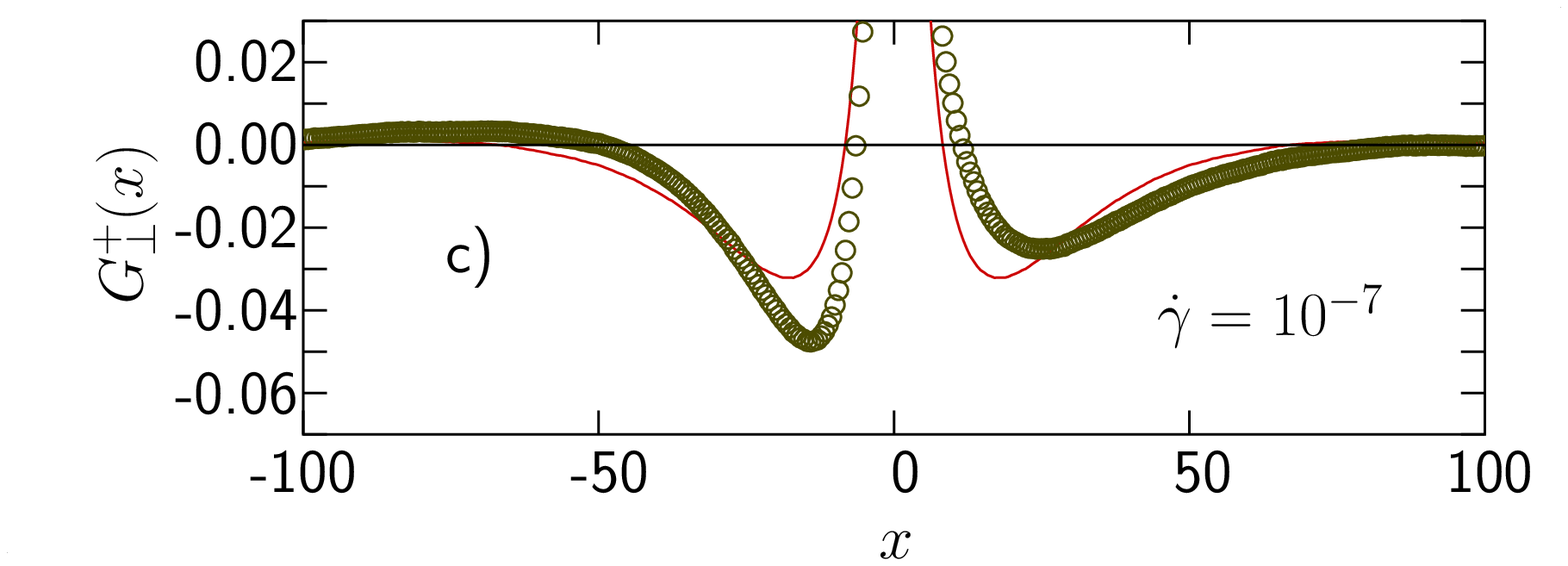}
  \includegraphics[width=8cm]{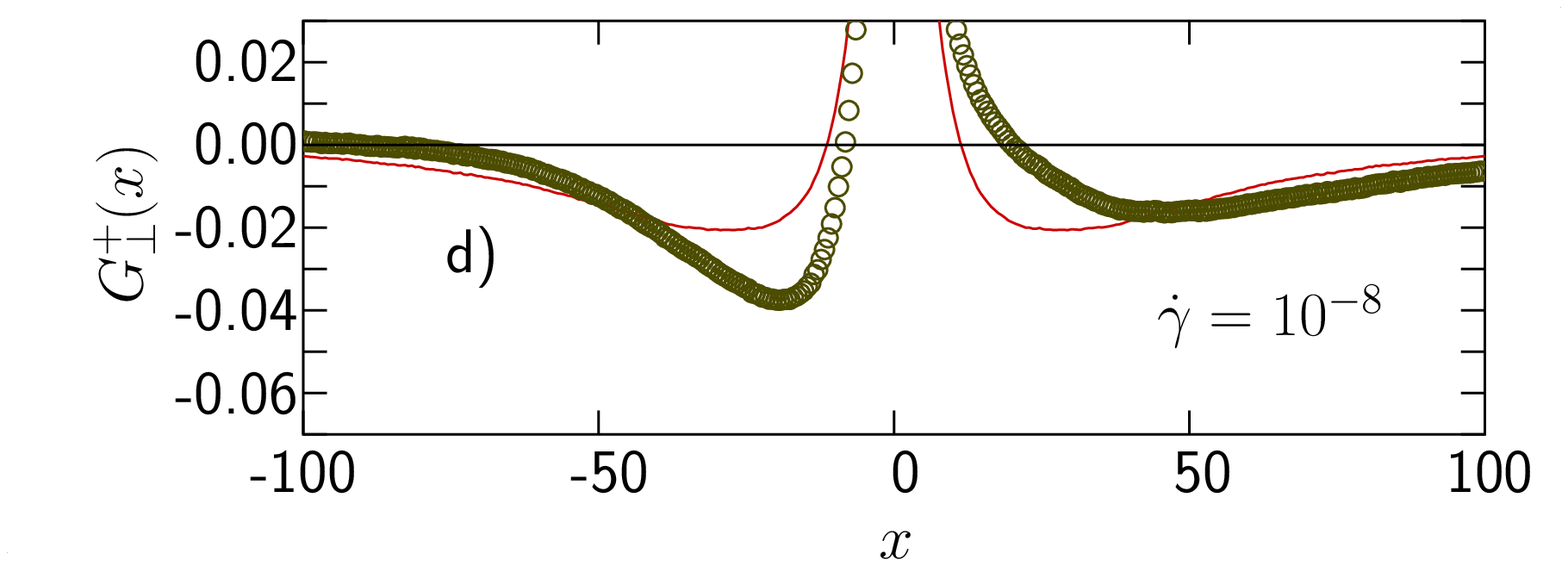}
  \caption{The correlation function $\Gy^+(x)$ at $\phi=0.8433\approx\phiJ$ and
    several different shear rates down to $\gdot=10^{-8}$. The asymmetry
    increases slowly with decreasing shear rate.}
  \label{fig:Gy}
\end{figure}

\begin{figure}
  \includegraphics[width=8cm]{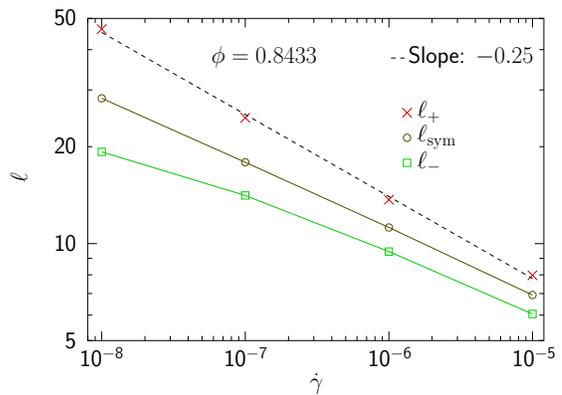}
  \caption{Position of the minima of $\gy^+(x)$ and $\gy$, respectively, at
    $\phi=0.8433\approx \phiJ$ and four different shear rates. Note the
    algebraic increase of $\ell_+\sim\gdot^{-0.25}$ which would imply a
    divergence at vanishing $\gdot$.}
  \label{fig:ell-gdot}
\end{figure}

\Fig{Gy} is the same quantity obtained at
$\phi=0.8433\approx\phiJ$ \cite{Heussinger_Barrat:2009} with
different shear rates. It is here found that the asymmetry grows with
decreasing shear rate. The algebraic increase of $\ell_+$, which
suggest a divergence in the limit of vanishing shear rate, is clear
from \Fig{ell-gdot} whereas both $\ell_-$ and $\ell_\sym$ grow
less rapidly.

\subsection{Origin of the rotational asymmetry}

\begin{figure}
  \includegraphics[width=8cm]{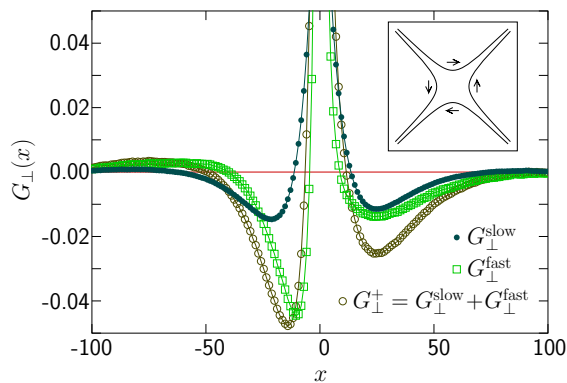}
  \caption{Splitting of the correlation function into two different
    contributions at $\phi=0.8433$ and $\gdot=10^{-7}$. The total $\Gy^+(x)$
    (open circles) is split into contributions from slow and fast particles,
    respectively.  It is clear that the asymmetry is largely an effect of the
    fast particles which we relate to the plastic processes. The inset shows an
    idealized velocity profile for a typical plastic event with quadrupolar
    structure. Note that this velocity field is a consequence of our shearing
    geometry which is equivalent to a compression along the $\pm(x-y)$
    directions together with an dilation in the orthogonal $\pm(x+y)$
    directions.}
  \label{fig:Gy-fast}
\end{figure}

We now turn to the question of the origin of the asymmetry in $\gy^+(x)$ and
will argue that it is linked to the plastic processes. The reason for this is
the manifest asymmetry of the velocity profile of an elementary plastic event,
as shown in the inset of \Fig{Gy-fast} \cite{Maloney_Lemaitre:2004}. Note that
the orientation of this velocity profile with its quadrupolar structure is
dictated by the direction of the shear. The velocity field corresponds to a
compression along the $\pm(x-y)$ direction together with an expansion in the
orthogonal $\pm(x+y)$ directions, which is equivalent to a simple shear. This
implies that the mirror version of such an event is expected to be much less
common, if at all present, and it is this effect that causes the asymmetry.
Note that this agrees well with the enhanced anti-correlation at $x<0$. When a
particle has $v_y>0$ due to a plastic event one would expect that there should
be one or more other particles with $v_y<0$, and the inset of \Fig{Gy-fast} shows
that they should be found in the $-x$ direction.

To further check this idea we have tried to separate the contribution to the
correlation function from the plastic events from the rest, i.e.\ we have
calculated $\gy^+$ separately for increasing and decreasing total energy
\cite{Ono_Tewari_Langer_Liu}, respectively. That study, which was done for
$N=4096$ particles, did indeed give evidence (not shown) that the part related
to a \emph{decrease} in total energy was the more asymmetric one. However, since
the total energy is a global quantity that kind of analysis isn't a very
sensitive one, at least not at finite shear rates. In a large system one would
expect some regions to be characterized by plastic events and a local decrease
in energy whereas the motion in other regions is elastic, but we still have to
classify the whole system as either plastic or elastic.

Instead of splitting up the contributions based on the change in the total
energy we now consider the change in the \emph{local} energy. That is done by
classifying each term in \Eq{gy-x} as ``fast'' or ``slow'', corresponding to
plastic and elastic respectively.  Since the change in total energy is $\Delta E
= L^2\sigma\gdot - \sum_i \v_i^2$ \cite{Ono_Tewari_Langer_Liu}, the change in
the local energy is related to the average $\v^2$ in a certain region. On the
average, one expects the energy to decrease locally if $\v^2>\expt{\v^2}$. As a
reasonable and more sensitive way to split $\gy^+(x)$ into two different
contributions, $\gy^+ = \gy^\slow + \gy^\fast$ (we drop the ``+'' in the
``slow'' and ``fast'' terms to simplify the notation) we therefore classify each
of the terms in \Eq{gy-x} according to the magnitude of the velocities. If both
$\v(\rp)$ and $\v(\rp + x\xhat)$ are ``slow'', (i.e.\ they both obey $\v^2 <
\expt{\v^2}$), the term contributes to $\gy^\slow$ whereas it contributes to
$\gy^\fast$ otherwise, i.e.\ if at least one of the particles is ``fast''.

\Figure{Gy-fast} shows the splitting of the total $\Gy^+(x)$ (open circles)
into slow and fast parts, respectively. Note that the contribution from the slow
particles (solid dots) is almost symmetric whereas there is a very pronounced
asymmetry from the fast particles (open squares). This is therefore evidence
that the asymmetry is caused by the fast particles and that this typically is
related to a local drop in energy which is often related to a plastic event.

\section{Summary}\label{sec:Summ}

To summarize, we have examined velocity correlations in a sheared system with
emphasis on the breaking of symmetries due to the shearing. We first find that
$\expt{v_x v_y}\neq0$ for individual particles and that this correlation depends
strongly on density. At low densities the particle motion is preferably
\emph{along} the force chains whereas it is preferably \emph{perpendicular to}
the force chains around $\phiJ$. We then examine how the total velocity
correlation $g(\rr)$ depends on the direction of $\rr$. Rather surprisingly the
correlation along the two diagonals, corresponding to the direction of
compression and dilation respectively, are almost identical. The decay along the
diagonals is non-monotonic, in contrast to the monotonic decay along the main
($\hat x$ and $\hat y$) directions.

We then argue that the usual correlation functions are more symmetric than the
system itself and define a less symmetric velocity correlation function that
also may be used to probe the differences between clockwise and
counter-clockwise rotations. This function is asymmetric with respect to $x$,
and this is an asymmetry that increases rather dramatically when either
the density increases above $\phiJ$ or the shear rate decreases at fixed
$\phi\approx\phiJ$. We attribute the asymmetry to elementary plastic events
with quadrupolar symmetry and their orientation dictated by the direction of the
shear as shown in the inset of \Fig{Gy-fast}.


I thank S. Teitel for helpful discussions and critical reading of an earlier
version of the manuscript. This work was supported by the Swedish Research
Council and the Swedish National Infrastructure for Computing.

%

\end{document}